\documentclass[letter]{aa} % for the letters 
\usepackage{graphicx}
\usepackage{txfonts}
\usepackage{hyperref}
\begin{document}

   \title{Observational Evidence of Velocity Anisotropy Assembly Bias in Galaxy Clusters}
 \author{Facundo Rodriguez\thanks{facundo.rodriguez@unc.edu.ar}
   \inst{1,2}
    \and
          Andrea Biviano\inst{3,4}
    \and 
          Ravi K. Sheth \inst{5} 
    \and
          Antonio D. Montero-Dorta\inst{6}
    \and
         Manuel Merchán \inst{1,2}
          }
   \institute{CONICET. Instituto de Astronomía Teórica y Experimental (IATE). Laprida 854, Córdoba X5000BGR, Argentina.
        \and
        Universidad Nacional de Córdoba (UNC). Observatorio Astronómico de Córdoba (OAC). Laprida 854, Córdoba X5000BGR, Argentina.
        \and 
        INAF-Osservatorio Astronomico di Trieste, Via G.B. Tiepolo, 11, I-34143 Trieste, Italy
        \and
        IFPU Institute for Fundamental Physics of the Universe, Via Beirut, 2, I-34014 Trieste, Italy
        \and
        Centre for Particle Cosmology, University of Pennsylvania, Philadelphia, PA 19104, USA
        \and
        Departamento de F\'isica, Universidad T\'ecnica Federico Santa Mar\'ia, Avenida Vicu\~na Mackenna 3939, San Joaqu\'in, Santiago, Chile.
             }
   \date{Received August 13, 2026; Accepted September 21, 2026}
   
% \abstract{}{}{}{}{} 
% 5 {} token are mandatory
 
  \abstract
   {Cosmological simulations show that halo clustering depends on secondary internal properties beyond mass, an effect known as assembly bias. In particular, numerical models have shown that internal velocity anisotropy ($\beta$) is the halo property most tightly linked to secondary halo bias: at fixed mass, haloes with isotropic velocity profiles are more strongly clustered, whereas haloes in which radial motions dominate are more weakly clustered.}
   {Despite this robust theoretical prediction, direct observational detection of velocity anisotropy assembly bias has remained elusive. Here we present observational evidence for this effect using a sample of massive galaxy clusters from the eROSITA and DESI surveys.}
   {We measure the internal orbital anisotropy $\beta$ for 91 clusters via Jeans dynamical modelling with MAMPOSSt, considering three anisotropy profile models to ensure robustness. We divided the sample into two subpopulations: isotropic and radially anisotropic, verifying that they show no significant differences in halo mass. To quantify their large-scale clustering, we measured the projected cross-correlation function with central galaxies from the SDSS DR18. We then calculated the average relative bias \( b_{\text{relative}} \) over scales of 5 to 20 \( h^{-1} \) Mpc.}
   {We report evidence of
velocity anisotropy assembly bias: clusters with isotropic internal kinematics exhibit stronger clustering than radially dominated ones, a segregation that is unlikely to
be attributable to residual differences in cluster mass. This trend shows varying significance across three anisotropy profiles, from a mild tendency to a highly pronounced signal. Furthermore, the assembly bias remains distinct when restricting the sample to clusters consistently classified across all three models. }
   {}


   \maketitle
%
%________________________________________________________________

\section{Introduction}
\label{Introduction}
The large-scale clustering of dark matter haloes is known to depend primarily on halo mass, as predicted by analytic models of structure formation \citep{Kaiser1984, Mo1996, ShethTormen1999, Sheth2001}. Cosmological simulations, however, have shown that halo clustering exhibits additional dependences on halo properties at fixed mass, a phenomenon commonly known as halo assembly bias \citep{Sheth2004, Gao2005, Wechsler2006}. These secondary dependences have been detected for a wide variety of halo properties, including formation time, concentration, spin, shape, and internal velocity structure \citep{Gao2007, Angulo2008, Faltenbacher2010, Lazeyras2017, Ramakrishnan2019, SatoPolito2019, Han2019}. Among these effects, the connection between halo clustering and internal dynamics at fixed halo mass is particularly intriguing. Using the Millennium Simulation, \cite{Faltenbacher2010} showed that velocity anisotropy is the halo property most tightly linked to halo bias, among those directly connected to the dynamical structure of haloes. Highly clustered haloes exhibit more isotropic internal motions systematically, while weakly clustered haloes are characterised by stronger radial anisotropy. Moreover, they found that this behaviour is common to several dynamical properties, suggesting that velocity anisotropy may encode a fundamental dynamical link between halo assembly history and large-scale environment.

Observationally, establishing the existence of halo assembly bias has proven considerably more challenging,  largely because halo properties are difficult to measure directly. Several studies have reported clustering dependences on galaxy colour, stellar age, and other galaxy properties at fixed halo mass \citep{Wang2008, Lacerna2014, MonteroDorta2017, Niemiec2018, Rodriguez2026}. While these trends do not constitute direct evidence of halo assembly bias, they may be related to it through the connection between central galaxy properties and the assembly history of their host haloes. Other works, however, have found little or no evidence for significant secondary clustering trends once halo mass is controlled for \citep{Lin2016, Sunayama2022}. A few observational studies have further extended these analyses to environmental and cosmic-web properties \citep{Paranjape2018, Alam2019, Rodriguez2026}, providing indirect insights into assembly-related effects. Direct observational constraints based on halo properties themselves remain scarce; one notable exception is the work of \citet{Kim2025}, who reported indications of a spin-dependent secondary bias. Nevertheless, the prediction of \citet{Faltenbacher2010} that halo clustering should depend directly on the internal orbital anisotropy of haloes has remained essentially untested in observational data.

Galaxy orbits in a cluster are expected to broadly trace the gravitational potential shaped by the dark matter distribution \citep[e.g.,][]{Biviano2004, Wojtak2009}. This is supported by $N$-body simulations: when subhaloes are selected to reproduce the observed spatial distribution of cluster galaxies, their velocity distribution and orbital anisotropy closely match those of the underlying dark matter
\citep{Faltenbacher2006}. Thus, it is reasonable to treat the anisotropy profile $\beta(r)$ derived from cluster members as an observational proxy for the internal orbital structure of the host halo, and we will do so in what follows. We caution, however, that this correspondence has so far been established in individual simulated haloes, and an exact quantitative equivalence between galaxy and dark matter anisotropy at the population level is not guaranteed.
Nevertheless, \cite{Biviano+26} have recently noted that, at fixed mass, more concentrated clusters have more radially anistropic motions.  While this is qualitatively consistent with the expected correlation for dark matter, the sample was too small to see if this imprints an assembly bias signal in the clustering. 

In this Letter, we present observational evidence of velocity anisotropy assembly bias using a sample of galaxy clusters drawn from the eROSITA and DESI surveys. We estimate the orbital anisotropy parameter $\beta$ through Jeans dynamical modelling using \texttt{MAMPOSSt} \citep{Mamon2013}, considering three anisotropy profiles to ensure robustness: a constant model ($\beta_C$), a radially increasing model ($\beta_T$), and an Osipkov-Merritt model ($\beta_{OM}$). We investigate whether the large-scale clustering of clusters depends on their internal velocity structure by measuring the projected cross-correlation function between clusters and central galaxies from the Sloan Digital Sky Survey (SDSS DR18) and computing the corresponding relative bias. We find that clusters with more isotropic internal motions are systematically more clustered than those dominated by radial orbits at similar mass, qualitatively consistent with theoretical expectations from cosmological simulations. Throughout this work, we adopt the standard $\Lambda$CDM cosmology \citep{Planck2016}, with parameters $\Omega_m = 0.3089$, $\Omega_b = 0.0486$, $\Omega_\Lambda = 0.6911$, and $H_0 = 100\, h \, \rm km \, \rm s^{-1} \, \rm Mpc^{-1}$, where $h = 0.6774$, $\sigma_8 = 0.8159$, and $n_s = 0.9667$. Throughout
this work, radii associated with the internal dynamical structure of
individual clusters (e.g. $r_{200}$, $r_\nu$, $r_s$, $r_{OM}$) are
expressed in physical (proper) Mpc at the cluster redshift, whereas
large-scale clustering separations ($r_p$, $w_p(r_p)$, and the scales
entering $\langle b_{\rm rel} \rangle$) are expressed in comoving
$h^{-1}\,{\rm Mpc}$, following standard convention in the respective
literatures.

%__________________________________________________________________

\section{Data}
\label{Data}

\subsection{Central galaxy sample}
\label{sub:central}

We construct a reference sample of central galaxies from the Sloan Digital Sky Survey (SDSS) Main Galaxy Sample \citep[MGS;][]{Strauss2002}, using the data from Data Release 18 (DR18; Almeida et al. 2023). 
We select all galaxies with spectroscopic redshifts $z < 0.3$ and $r$-band apparent magnitudes $r < 17.77$ to ensure high completeness. Galaxy groups are identified using the \citet{rodriguez2020} group finder, as implemented in the extended catalogue of \citet{Rodriguez2026}, where central galaxies are defined as the brightest group members. Our final sample comprises $441\,208$ central galaxies. This sample serves as a dense tracer of the large-scale density field for cross-correlation with the cluster sample. Using central galaxies instead of the full galaxy population minimises satellite contamination, providing a cleaner tracer of the halo bias on large scales.Tests on mock catalogues by \citet{rodriguez2020} show that this group finder
achieves an average group purity above 0.8 and a completeness above 0.9,
and the resulting central and satellite classification has since been
validated against independent weak-lensing mass estimates and cosmological simulations \citep[e.g.:][]{Gonzalez2021, rodriguez2021, RodriguezMedrano2023, alfaro2022, Izzo2026}.

\subsection{Cluster sample}
\label{Cluster sample}
Initially, we select 374 clusters from the eROSITA catalogues of \citet{Bulbul+24} and \citet{Sanders+25} with the following properties, 
%\begin{itemize}
(i) mass $M_{500,X}\footnote{$M_{\Delta} = 4\pi\,\rho_c(z)\, r^3_\Delta/3 $ is the mass at the radius $r_\Delta$ enclosing a spherical overdensity $\Delta$ times the critical density of the Universe $\rho_c(z)$ at redshift $z$. We use the suffix 'X' to indicate that the masses are estimated from X-ray data.} \geq 10^{14} \, M_{\odot}$,
(ii) extent likelihood ${\cal L}_{\mathrm{ext}} > 10$,
and (iii) centre coordinates inside two high-spectroscopic completeness regions of the 1$^{st}$ data release of the Dark Energy Spectroscopic Instrument \citep[DESI DR1;][]{DESIDR1}, namely $(120^{\circ} \leq \mathrm{RA} \leq 260^{\circ}) \cap (\delta \leq 10^{\circ})$ and  $(180^{\circ} \leq \mathrm{RA} \leq 250^{\circ}) \cap (30^{\circ} \leq \delta \leq 40^{\circ})$. 
%\end{itemize}

We then apply the \texttt{Clean} algorithm for cluster member identifications \citep{Mamon2013} to these clusters, and we retain 192 of them, discarding those with fewer than 10 members within $r_{200}$. Of these 192, we finally restrict our analysis to the 102 clusters that are also present in the SDSS group catalogue (see Sect.~\ref{sub:central}). 

%-------------------------------------------------------------------
\section{Methods}
\label{Methods}
\subsection{Velocity anisotropy measurements}
\label{Velocity anisotropy measurements}
Given the cluster $M_{500,X}$ and redshift values in the catalogue of \citet{Bulbul+24}, we evaluate $r_{200,X}$ by adopting a NFW mass profile \citep{NFW96} with concentrations, $c_{500}$\footnote{$c_{500} \equiv r_{500}/r_s$, where $r_s$ is the scale radius of the NFW profile.} evaluated from $M_{500,X}$ using the $z$-dependent $c_{500}-M_{500}$ relation of \citet{Ragagnin+21}. We restrict our dynamical analysis to the galaxies in the radial range 0.05 Mpc -- $r_{200,X}$, to exclude the central region usually dominated by the gravitational potential of the brightest cluster galaxy, and the external regions, dominated by the infalling, unvirialized galaxy population \citep[see, e.g.,][]{Biviano+13}.

We determine the projected number density profiles of each cluster, $N(R)$, using only cluster members, by adopting the cluster centres in the catalogue of \citet{Bulbul+24}, and correcting for incompleteness due to fibre collision (see Appendix~\ref{app:compl}). To ensure that our results do not depend on the fibre collision correction, we also consider the case in which the sample is assumed to be complete. As we describe in Appendix~\ref{app:dyna}, our results are robust with respect to the incompleteness correction.

We fit a NFW model \citep[in projection,][]{Bartelmann96} to $N(R)$, using the maximum likelihood technique described by \cite{Sarazin80}, to find the best-fit value of the $r_{\nu}$ scale radius parameter of each cluster. The procedure did not converge for 11 of the 102 clusters, suggesting there could be a problem with the X-ray centre identification. In the following we restrict our analysis to the 91 remaining clusters. Our final sample of 91 clusters spans the range $0.05 \leq z \leq 0.30$, with a median $z=0.15$, and contains from 11 to 241 cluster members within $r_{200}$, with a median of 50.

We then run
%\footnote{More details on the dynamical procedure can be found in Appendix~\ref{app:dyna}.}
the \texttt{MAMPOSSt} code of \citet{Mamon2013,Pizzuti+23} on each of these 91 clusters to determine its dynamical properties by application of the Jeans equation.  We adopt the NFW model for the mass density profile $\rho(r)$, characterised by the two parameters $r_{200}, r_s$, where $r_s$ is the scale radius of the NFW model. Note that the NFW profiles describing the galaxy and the mass distributions are not the same, and $r_{\nu}$ and $r_s$ are different parameters.

For the sake of generality, we consider three different models for the velocity anisotropy profile, $\beta(r)$,
\begin{itemize}
\item \texttt{constant}: $\beta_C(r) = C$; 
\item \texttt{Tiret} \citep{Tiret+07}: $\beta_T(r) = \beta_{\infty} \, r/(r+r_s)$; \item \texttt{OM} \citep{Osipkov79,Merritt85}: $\beta_{OM}(r) = r^2/(r^2+r_{OM}^2)$,
\end{itemize}
where $\beta \equiv 1-\sigma_{t}^2/ \sigma_r^2$, $\sigma_t$ and $\sigma_r$ are the tangential, and, respectively, radial component and of the velocity dispersion tensor, and where we assume the two tangential components to be identical, $\sigma_{\theta} = \sigma_{\phi} \equiv \sigma_t$. Each of the three models is characterised by a single anisotropy parameter. Higher values of $C$ and $\beta_{\infty}$, and lower values of $r_{OM}$, indicate more radially anisotropic orbits of cluster galaxies. All models assume $\beta(0)=0$ (central isotropic orbits), an assumption supported by independent studies of $\beta(r)$ in nearby clusters \citep{WL10,Li+23}.
Including additional parameters in our dynamical analysis is not recommended given the typical number of cluster members available in our sample. 

\subsection{Measurement of relative bias}
\label{sec:Measurement_of_relative_bias}

Following \citet{Rodriguez2026}, for each cluster subsample we compute the
projected cross-correlation function $w_{\mathrm{p}}(r_p)$ with the central
galaxy sample using the Landy--Szalay estimator, integrating over the
line-of-sight separation from $0.1$ to $\pi_{\max}=50\,h^{-1}\mathrm{Mpc}$.
The large-scale relative bias, $\langle b_{\rm rel}\rangle$, is obtained by
averaging over comoving scales between $5$ and $20\,h^{-1}\mathrm{Mpc}$, with
uncertainties from a joint jackknife scheme that accounts for the covariance
between subsamples (Appendix~\ref{app:bias}).

%-------------------------------------------------------------------

\section{Results: Anisotropy-dependent clustering bias}
\label{sec:anisotropy_bias}

For each estimator ($C$, $\beta_\infty$, and $r_{\rm OM}$), we split the
sample into low- and high-anisotropy subpopulations using both median and
quartile splits (Table~\ref{tab:subsamples}). Note that large values of
$C$ and $\beta_\infty$ indicate radial orbits, whereas large values of
$r_{\rm OM}$ indicate isotropic ones. We also construct an intersection
sample of clusters classified consistently by all three median splits,
yielding 30 isotropic and 27 radially anisotropic clusters. The
anisotropy-based subsamples show no significant differences in mass
(Appendix~\ref{app:radius}), a necessary condition for interpreting
clustering differences as assembly bias.

\begin{figure}
\centering
\includegraphics[width=0.98\columnwidth]{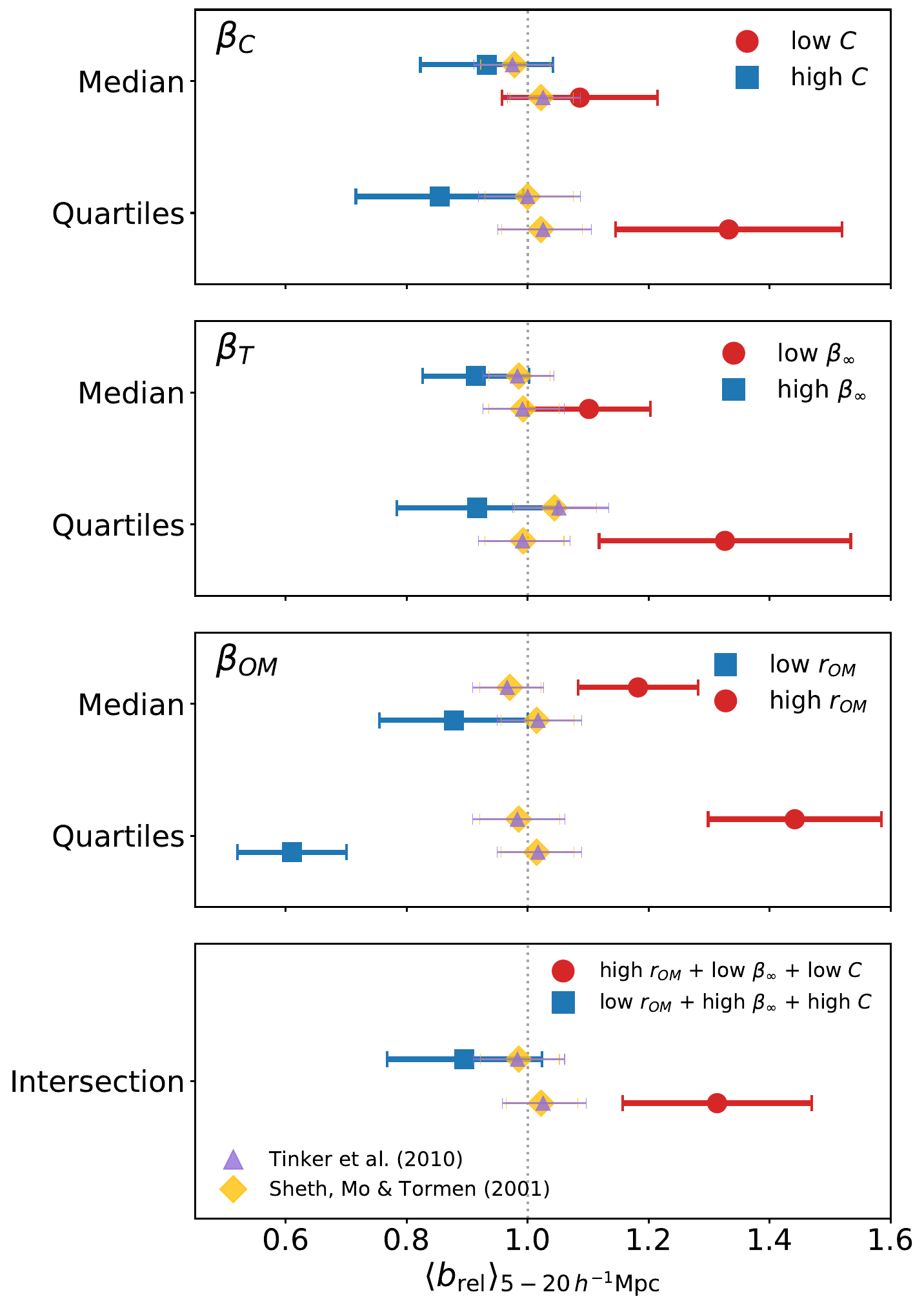}
\caption{Comparison between the observed large-scale relative bias and theoretical
predictions derived from mass-only halo bias models. The horizontal axis shows the average
relative bias $\langle b_{\mathrm{rel}} \rangle$ averaged over scales of
$5 < r_p < 20\,h^{-1}\mathrm{Mpc}$. Panels from top to bottom present results for the
$\beta_C$, $\beta_T$, and $\beta_{OM}$ models (median and quartile splits), followed by the
intersection sample. In all cases, the observed low-$\beta$ (isotropic or tangential) and
high-$\beta$ (radial) subpopulations are shown as red circles and blue squares, with
horizontal bars indicating the $1\sigma$ jackknife uncertainties. Purple triangles and amber diamonds denote the corresponding theoretical predictions computed using the halo bias models of \citet{Tinker2010} and \citet{Sheth2001}, respectively.}
\label{fig:brel_theory_comparison}
\end{figure}

Figure~\ref{fig:brel_theory_comparison} summarises the large-scale average relative bias values, $\langle b_{\rm rel}\rangle$, together with mass-only theoretical predictions from \citet{Tinker2010} and \citet{Sheth2001} integrated over the $r_{200}$ distributions of each subpopulation. In all cases, the observed mean relative bias for the isotropic subsystems (red circles) systematically exceeds the theoretical mass-only predictions, while the radially anisotropic ones (blue squares) fall below them. This dynamical segregation is particularly clear for the $\beta_{OM}$ model and the intersection sample (bottom panel). Notably, the assembly bias signal is most pronounced in the isotropic/tangential subpopulations, where the departure from the mass-only baseline is clearest. Conversely, the theoretical predictions (purple triangles and amber diamonds) remain tightly clustered near unity, suggesting that residual mass differences are unlikely to account for the observed splitting.

We present the scale-averaged bias difference $\Delta b_{\text{rel}}$, its covariance-aware uncertainty $\sigma_{\Delta}$, and the signal-to-noise significance $S_{\Delta}$ for all anisotropy profile models, sample splits, and the intersection sample in Table~\ref{tab:significance}.

For the quartile splits, the isotropic subpopulations systematically exhibit stronger clustering than the radially anisotropic ones across all three profile models. In particular, the $\beta_{\text{OM}}$ model yields the most pronounced segregation among individual estimators ($S_{\Delta} = 5.30$); remarkably, a generalised form of the $\beta_{\text{OM}}$ model was shown by \citet{Biviano+26} to provide a better fit than the generalised $\beta_T$ model for a sample of nine very well-sampled clusters, which aligns with its stronger performance in capturing velocity anisotropy variations. Meanwhile, $\beta_C$ and $\beta_T$ display moderate significance ($S_{\Delta} = 1.67$ and $1.51$, respectively). For the median splits, the signal remains consistently positive across all models with lower overall significance ($S_{\Delta} \sim 0.65 - 1.57$), as expected from the reduced dynamical contrast. Notably, the robust intersection sample yields a significant bias difference of $\Delta b_{\text{rel}} = 0.418 \pm 0.244$ ($S_{\Delta} = 1.71$). As a check against orientation-dependent triaxiality effects, we
verified that the observed segregation is stable to changes in the
line-of-sight integration limit $\pi_{\max}$ (Sect.~\ref{sec:Measurement_of_relative_bias}).

There exists the possibility that the mass--anisotropy degeneracy inherent to
the joint fit could induce the observed clustering segregation. To assess this,
we re-ran MG-MAMPOSSt fixing $r_{200}$ to the independent eROSITA value
$r_{200,\mathrm{X}}$. The resulting subsamples remain mass-matched
(Appendix~\ref{app:r200X}), and the isotropic clusters are still more strongly clustered
than the radially anisotropic ones (Appendix~\ref{app:degeneracy-free}).

\begin{table}[t]
\centering
\caption{Scale-averaged relative bias difference $\Delta b_{\text{rel}}$ ($5$--$20\,h^{-1}\text{Mpc}$), joint jackknife error $\sigma_{\Delta}$, and signal-to-noise ratio $S_{\Delta}$.}
\label{tab:significance}
\begin{tabular}{lcccc}
\hline\hline
Sample Split & Estimator & $\Delta b_{\text{rel}}$ & $\sigma_{\Delta}$ & $S_{\Delta}$ \\
\hline
Quartiles & $\beta_C$ & $0.478$ & $0.286$ & $1.67$ \\
 & $\beta_T$ & $0.410$ & $0.272$ & $1.51$ \\
 & $\beta_{\text{OM}}$ & $0.831$ & $0.157$ & $5.30$ \\
\hline
Median & $\beta_C$ & $0.154$ & $0.236$ & $0.65$ \\
 & $\beta_T$ & $0.187$ & $0.184$ & $1.02$ \\
 & $\beta_{\text{OM}}$ & $0.305$ & $0.194$ & $1.57$ \\
\hline
Intersection & All three & $0.418$ & $0.244$ & $1.71$ \\
\hline
\end{tabular}
\end{table}

%-------------------------------------------------------------------
\section{Conclusions}
\label{sec:Conclusions}
In this Letter, we have presented observational evidence of velocity anisotropy assembly bias in galaxy clusters, as traced by the orbital anisotropy of cluster galaxies, using a sample of massive galaxy clusters from the eROSITA and DESI surveys. By partitioning the sample into isotropic and radially anisotropic subpopulations across three different internal kinematic models ($\beta_C$, $\beta_T$, and $\beta_{OM}$), we find a systematic difference in their large-scale clustering amplitude. For samples with similar halo mass distribution (proxied by $r_{200}$), clusters dominated by isotropic internal motions consistently display a higher relative bias than those dominated by radial orbits, suggestive of a dynamical link between the internal orbital structure of clusters and their cosmic web environment.

While this signature is consistently found across models and selection criteria in our current dataset, the statistical precision remains constrained by our limited sample size, particularly for the isotropic subpopulations. Looking forward, this analysis establishes a timely baseline for upcoming deep spectroscopic surveys. Future data releases from instruments like DESI will vastly expand cluster samples with denser spectroscopic coverage. These imminent data streams will allow higher-precision orbital measurements, enabling finer mass stratifications and firmly establishing velocity anisotropy as a key observational tracer of halo assembly history.

\begin{acknowledgements}
      {We thank an anonymous referee for her/his useful remarks and suggestions that helped improve the robustness of our results.}
\end{acknowledgements}

%-------------------------------------------------------------------

% for the bibliography, at the end
\bibliographystyle{aa} % style aa.bst
\bibliography{main} % your references Yourfile.bib

\begin{appendix}

\section{Details of the velocity anisotropy measurement procedure}
\subsection{Correction for spectroscopic incompleteness}
\label{app:compl}
The DESI spectroscopic survey suffers from incompleteness due to fibre assignment \citep{Bianchi+25}. This incompleteness affects our analysis because it imprints a selection function that depends on the inter-galaxy angular distance, and must be corrected to retrieve the intrinsic spatial distribution of cluster galaxies. We use the function
\begin{equation}
C = 0.47 + 0.53 \, (1+10^{-2.75 ( \log(\theta)+1.85)})^{-1}
\end{equation}
where $\theta$ is the inter-galaxy angular distance in degrees, to fit the incompleteness curve in Fig.~3 (middle panel) of \citet{Bianchi+25}. We then apply a 
$1/C(\theta_i)$ weight to each galaxy in our maximum likelihood fitting procedure of $N(R)$ by the NFW profile (in projection). Note that these weights do not enter the dynamical analysis other than through the estimate of the number density profile $N(R)$. In fact, spectral incompleteness is not expected to affect the distribution of galaxy velocities, since from an observational point of view it is impossible to bias the selection of galaxies in velocity space within the restricted redshift range covered by cluster galaxies.

In the first iteration of the fitting procedure we use the observed $\theta_i$. We then correct these values to account for the fact that the observed inter-galaxy distances are larger than the original ones before the fibre assignment. We re-evaluate $\theta_i$ from the galaxy surface density implied by the best-fit NFW profile \citep{LM01}. We then use the new $\theta_i$ to compute new weights in the fitting procedure, and iterate. We stop the iteration when $c_{200,\nu} \equiv r_{200}/r_{\nu}$ changes by $\leq 10$\% with respect to the previous iteration.

\subsection{Details on the dynamical analysis}
\label{app:dyna}
In our dynamical analysis, we solve the Jeans equation for dynamical equilibrium using the \texttt{MAMPOSSt} algorithm of \citet{Mamon2013}, in the \texttt{MG-MAMPOSSt} implementation of \citet{Pizzuti+23}, by restricting to the classical general relativity case.
In our \texttt{MG-MAMPOSSt} runs we adopt Gaussian priors for $r_{200}$ and $r_s$, different for each cluster. The priors for $r_{200}$ are centred on the $r_{200,X}$ values derived from the catalogue of \citet{Bulbul+24}, with $\sigma(r_{200})$ equal to the error on $r_{200,X}$ as obtained from the $M_{500,X}$ uncertainties via error propagation analysis. For $r_s$ we centre the priors on $r_{200,X}/4$, since 4 is a typical value for $c_{200}$ of low-$z$ massive clusters \citep[e.g.,][]{Ragagnin+21}, with $\sigma(r_s)=<(r_{200,X}-\sigma(r_{200,X}))/6,(r_{200,X}+\sigma(r_{200,X}))/2>$, that reflects the large uncertainty in the $r_s$ prior. We adopt flat, uninformative, priors on the parameters describing the velocity anisotropy profiles, $\beta_C, \beta_T,$ and $r_{OM}$.

The $r_{\nu}$ values are taken from the $N(R)$ fitting procedure, that accounts for incompleteness in the spectroscopic sample. To assess the importance of the incompleteness correction on our results, we also run \texttt{MG-MAMPOSSt} on the $r_{\nu}$ values we obtained by the same $N(R)$ fitting procedure, without correction. We find that, despite a significance change in the $r_{\nu}$ values ($\Delta r_{\nu}/r_{\nu} = 2 \pm 3$), the values of $r_{200}$ we found are almost unaffected ($\Delta r_{200}/r_{200}$ vary between -0.01 and -0.02, with r.m.s. 0.03-0.04, depending on the $\beta$ model), and also the values of $\beta$ are not significantly modified ($\Delta \beta \simeq -0.08$ with r.m.s between 0.17 and 0.23 for the T and C model, respectively, and $\Delta r_{OM}=0.3 \pm 0.4$ Mpc). 

We run \texttt{MG-MAMPOSSt} using a Monte Carlo Markov Chain (MCMC) procedure with 15,000 steps, using the \citet{GR92} criterion to check for convergence, adopting a threshold of $\hat{R}=1.01$ for the Gelman-Rubin coefficient. This procedure is repeated three times for each cluster, once for each $\beta(r)$ model.

To determine whether the three adopted anisotropy models yield consistent descriptions of the internal kinematics, we examine the correlations among their best-fit parameters. We find a strong positive correlation between the $C$ and $\beta_{\infty}$ best-fit values resulting from our dynamical analysis, with a Spearman rank correlation coefficient \citep[see, e.g.,][]{Press2007Numerical} $\rho_S=0.78$, corresponding to a probability $p<0.001$. The best-fit values of $r_{OM}$ show very significant anti-correlations with both $C$ ($\rho_S=-0.62, p<0.001$) and $\beta_{\infty}$ ($\rho_S=-0.76, p<0.001$).

Clusters are known to be preferentially prolate \citep{SDFLB06}, and projected orbital anisotropy is degenerate with the orientation of their main axis with respect to the line-of-sight \citep{Wojtak+13}. We do not have access to the 3D structure of clusters in our sample, but we can use the ellipticity of their X-ray emission \citep[from][]{Sanders+25} as a proxy. We find no correlation between our 91 cluster ellipticities and any of their velocity anisotropy parameter. This suggests that our estimates of velocity anisotropies are not biased by the orientation of the cluster along the line-of-sight.

\section{Relative bias estimation procedure}
\label{app:bias}
The large-scale relative bias is defined as the ratio over the full sample:
\begin{equation}
b_{\mathrm{rel}}^{S} = \left\langle \frac{w_{\mathrm{p}}^{Sub}(r_p)}{w_{\mathrm{p}}^{\mathrm{total}}(r_p)} \right\rangle_{5-20\,h^{-1}\mathrm{Mpc}},
\end{equation}
averaged over comoving scales $5 < r_p < 20\,h^{-1}\mathrm{Mpc}$ where the bias is scale-independent.
The cross-correlation is computed against $5\times10^{7}$ random points
matching the survey mask and the angular and redshift selection of the
central-galaxy sample (Sect.~\ref{sub:central}); no additional pair weights
are applied. Uncertainties are estimated via a stratified jackknife scheme,
with $N_{\rm JK} = \min(n_{\rm low}, n_{\rm high})$ groups per split (23--45
depending on the split), constructed so that every realisation removes a
balanced number of clusters from both the low- and high-anisotropy
subsamples simultaneously. This yields a direct estimate of the
cross-covariance between the two clustering measurements, rather than
sampling large-scale spatial variance. We verified that our results are
stable to reasonable changes in the minimum scale of the averaging window
and in $\pi_{\max}$.

To quantify the clustering segregation between the isotropic and radially anisotropic subpopulations while fully accounting for their mutual cross-covariance, we define the difference in scale-averaged relative bias as our primary statistic:
\begin{equation}
\mathbf{\Delta b_{\text{rel}} \equiv \langle b_{\text{rel}}^{\text{iso}} \rangle_{5-20\,h^{-1}\text{Mpc}} - \langle b_{\text{rel}}^{\text{rad}} \rangle_{5-20\,h^{-1}\text{Mpc}}.}
\end{equation}
Because both subpopulations share the same reference sample of central galaxies and full-cluster normalisation, their relative bias measurements are intrinsically correlated.

To evaluate the uncertainty $\sigma_{\Delta}$ on this difference, all projected cross-correlation functions and their corresponding relative bias ratios are recomputed jointly within each jackknife realisation. The variance of the primary statistic is explicitly evaluated as:
\begin{equation}
\mathbf{\text{Var}(\Delta b_{\text{rel}}) = \text{Var}(\langle b_{\text{rel}}^{\text{iso}} \rangle) + \text{Var}(\langle b_{\text{rel}}^{\text{rad}} \rangle) - 2\,\text{Cov}(\langle b_{\text{rel}}^{\text{iso}} \rangle, \langle b_{\text{rel}}^{\text{rad}} \rangle),}
\end{equation}
where the cross-covariance term $\text{Cov}(\langle b_{\text{rel}}^{\text{iso}} \rangle, \langle b_{\text{rel}}^{\text{rad}} \rangle)$ is determined directly across the joint jackknife realisations. The error on the bias difference is defined as $\sigma_{\Delta} \equiv \sqrt{\text{Var}(\Delta b_{\text{rel}})}$, and the associated signal-to-noise significance is given by $S_{\Delta} \equiv \Delta b_{\text{rel}} / \sigma_{\Delta}$.

\section{Systematic differences in cluster mass}
\label{app:mass}

\subsection{Mass consistency from the dynamical MG-MAMPOSSt r200}
\label{app:radius}

A necessary validation step before interpreting the clustering measurements is to verify that our anisotropy-based subpopulations are not segregated by halo mass. Given the limited size of our sample, we cannot perform a fine binning in mass; instead, we adopt the characteristic radius $r_{200}$ derived from the MG-MAMPOSSt dynamical analysis as a proxy for cluster mass, and compare its distribution across the isotropic and radial subsamples defined in Table~\ref{tab:subsamples}.

\begin{table}[h]
\centering
\caption{Velocity anisotropy thresholds used to define the cluster subpopulations.}
\label{tab:subsamples}
\begin{tabular}{lcccc}
\hline\hline
Model (parameter) & \multicolumn{2}{c}{Quartile Split} & & Median Split \\
\cline{2-3} \cline{5-5}
 & Low & High & & Threshold \\
\hline
$\beta_C$ ($C$) & -0.23 & 0.62 & & 0.41 \\
$\beta_T$ ($\beta_{\infty}$) & -0.04 & 0.69 & & 0.49 \\
$\beta_{OM}$ ($r_{OM}$ [Mpc]) & 1.45 & 2.70 & & 2.10 \\
Intersection &  - & - & & All three estimators\\
\hline
\end{tabular}
\tablefoot{The table lists the physical values of the anisotropy parameter in the $\beta(r)$ models, used as boundaries for each selection method. For the intersection sample, clusters are classified as isotropic (radially anisotropic) if they fall below (above) the median threshold in \(C\) and \(\beta_\infty\), and above (below) the median threshold in \(r_{OM}\), simultaneously for all three estimators.}
\end{table}

\begin{figure}
\centering
\includegraphics[width=0.98\columnwidth]{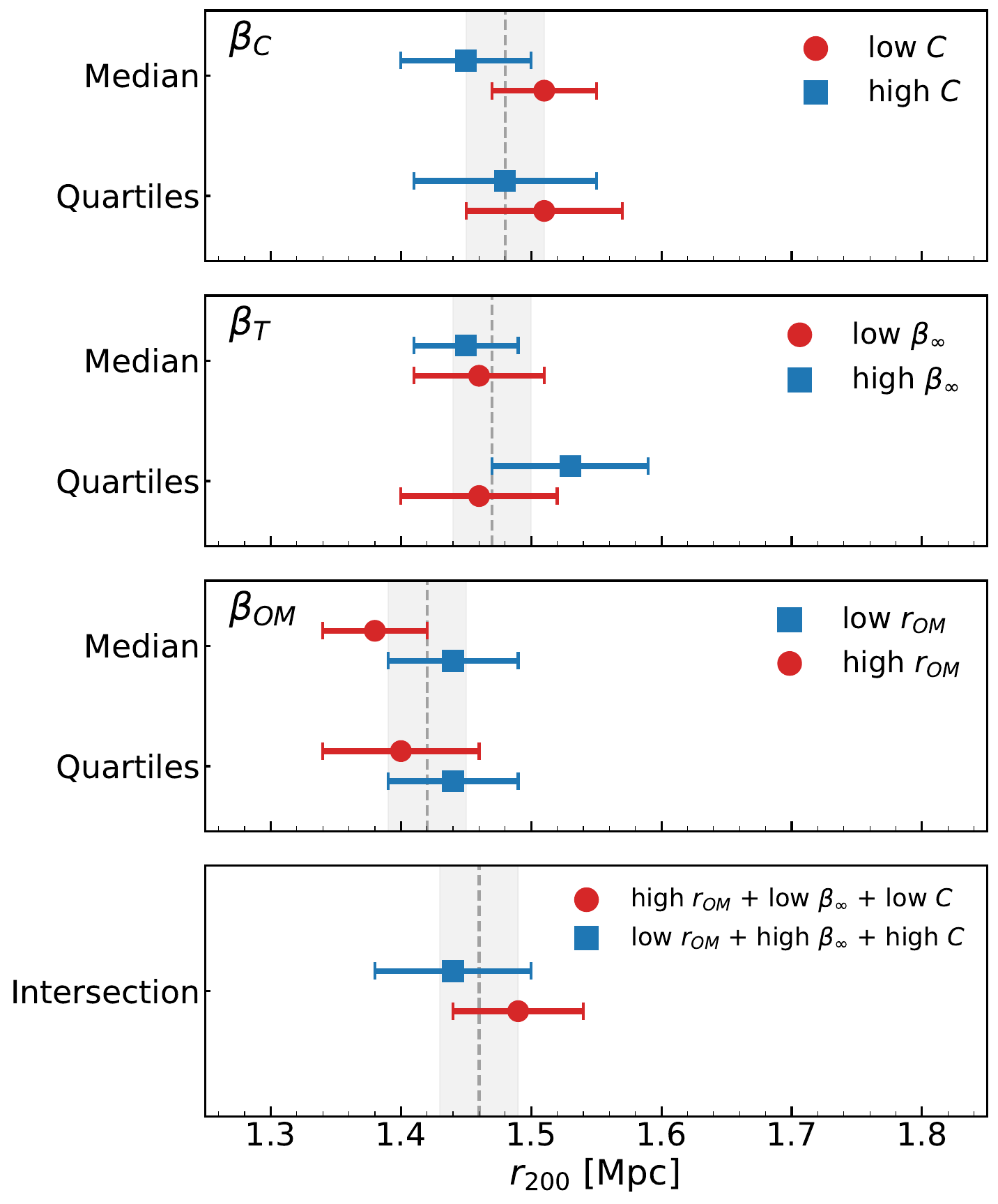}

\caption{Cluster radius $r_{200}$ measurements for subpopulations split by velocity anisotropy. The top three panels correspond to individual anisotropy estimators ($\beta_C$, $\beta_T$, and $\beta_{OM}$) evaluated via median and quartile splits. The bottom panel shows the robust intersection sample, for which the $r_{200}$ values are obtained by averaging the radii measured with the three individual anisotropy methods. Red circles denote isotropic-dominated subsystems (low $\beta$), while blue squares represent radially dominated ones (high $\beta$). The vertical dashed line and shaded area indicate the total sample mean and its corresponding $1\sigma$ uncertainty. Error bars represent $1\sigma$ uncertainties for each subsample. \texttt{MG-MAMPOSSt} $r_{200}$ values are used, so they can be different for the same cluster sample when different $\beta(r)$ models are used. }
\label{fig:r200_comparison}
\end{figure}

Figure~\ref{fig:r200_comparison} presents the mean $r_{200}$ values for each subpopulation, along with the total sample mean and its $1\sigma$ uncertainty (dashed vertical line and shaded band). For all anisotropy models and selection criteria (median and quartile splits, as well as the intersection sample), the mean $r_{200}$ values of the low- and high-anisotropy subsamples are consistent within the $1\sigma$ uncertainties. A permutation test \citep[e.g.][]{good2005permutation} confirms that the differences are not statistically significant. Moreover, a Spearman rank correlation analysis shows no significant correlation between the individual anisotropy parameters ($C$, $\beta_\infty$, or $r_{\rm OM}$) and $r_{200}$. These tests indicate that our kinematic selection does not introduce significant mass biases in this dynamical mass proxy.

We note that the $r_{200}$ estimates shown in Fig.~\ref{fig:r200_comparison} are model-dependent, as they are obtained from MG-MAMPOSSt for each assumed anisotropy profile. This explains why, for a given split, the $r_{200}$ values can vary slightly between the $\beta_C$, $\beta_T$, and $\beta_{OM}$ panels.
Nevertheless, the consistency across panels reinforces the robustness of our conclusion: the subsequent clustering signal is unlikely to be
attributable to residual mass differences. Such residual mass differences are already accounted for in the main text through the mass-only theoretical predictions shown in Fig.~\ref{fig:brel_theory_comparison} (\citet{Tinker2010} and \citet{Sheth2001}, purple triangles and amber diamonds, respectively).

We note, however, that in the MG-MAMPOSSt fit $r_{200}$ is treated as a free parameter and is therefore constrained jointly with the anisotropy parameters. As a consequence, this dynamical $r_{200}$ is not fully independent of $\beta$, and the test above alone cannot completely rule out an entanglement between the mass–anisotropy degeneracy of the joint fit and the observed clustering segregation. To address this concern directly, in Appendices \ref{app:r200X} and \ref{app:degeneracy-free} we repeat the mass-consistency check and the clustering measurement using r200 fixed to the independent, degeneracy-free eROSITA value r200,X.

\subsection{Robustness to the mass–anisotropy degeneracy: fixing r200 to the independent eROSITA mass}
\label{app:r200X}

To explicitly eliminate any potential systematic arising from the mass--anisotropy degeneracy, we re-ran MAMPOSSt for all clusters by fixing $r_{200}$ directly to the independent X-ray radius $r_{\rm X} \equiv r_{200,\rm X}$ derived from eROSITA \citep{Bulbul+24}. Under this setup, the determination of the halo mass is completely decoupled from the galaxy velocity dispersion, yielding estimates of the velocity anisotropy parameters ($C$, $\beta_\infty$, and $r_{\rm OM}$) that are free from dynamical degeneracies with $r_{200}$.

Because fixing $r_{200} = r_{200,\rm X}$ modifies the posterior MAMPOSSt distributions, the best-fit anisotropy parameters shift slightly compared to the baseline run. Consequently, we recomputed the corresponding median and quartile threshold values for each model to partition the sample into new low- and high-anisotropy subpopulations. We then repeated the exact same mass-consistency analysis applied in the main text to these updated subsamples.

\begin{figure}
\centering
\includegraphics[width=0.98\columnwidth]{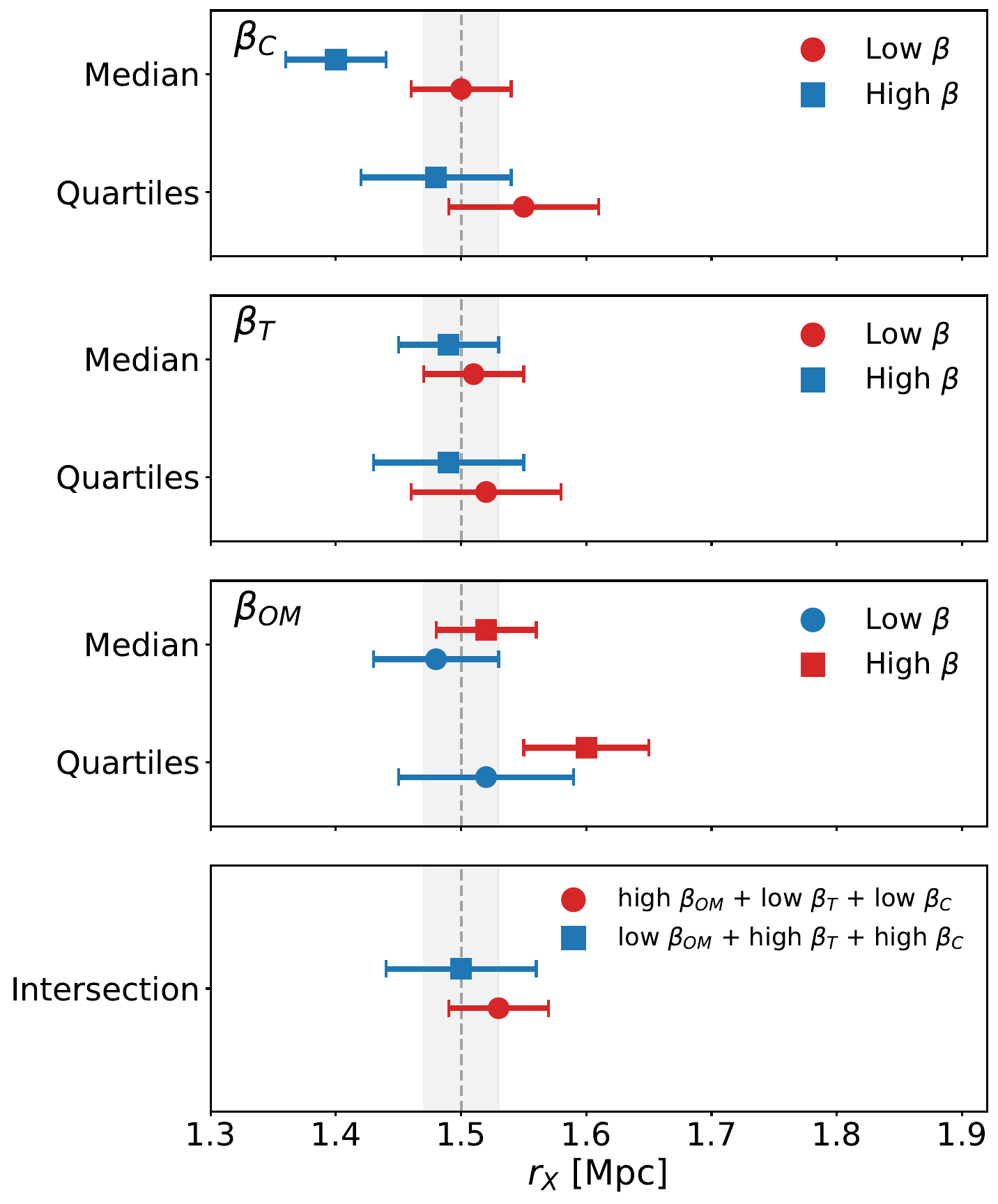}

\caption{Same as Fig.~\ref{fig:r200_comparison}, but showing the mean X-ray radius $r_{\rm X} \equiv r_{200,\rm X}$ for cluster subpopulations.}
\label{fig:rx_summary}
\end{figure}

Figure~\ref{fig:rx_summary} displays the mean values $r_{\rm X}$ for the updated low- and high-anisotropy subsamples across all profile models and selection criteria, alongside the complete sample mean and its uncertainty $1\sigma$ (dashed line and shaded band). As in the baseline case, the mean X-ray radii of the isotropic (low $\beta$) and radially anisotropic (high $\beta$) subpopulations are consistent within $1\sigma$ uncertainties across all splits.

\subsection{Clustering signal under a degeneracy-free mass proxy}
\label{app:degeneracy-free}

Having confirmed that fixing $r_{200} = r_{200,\rm X}$ yields clean, mass-matched subsamples free from dynamical degeneracies, we re-evaluate the large-scale clustering signal to verify whether the observed anisotropy assembly bias persists.

\begin{figure}
\centering
\includegraphics[width=0.98\columnwidth]{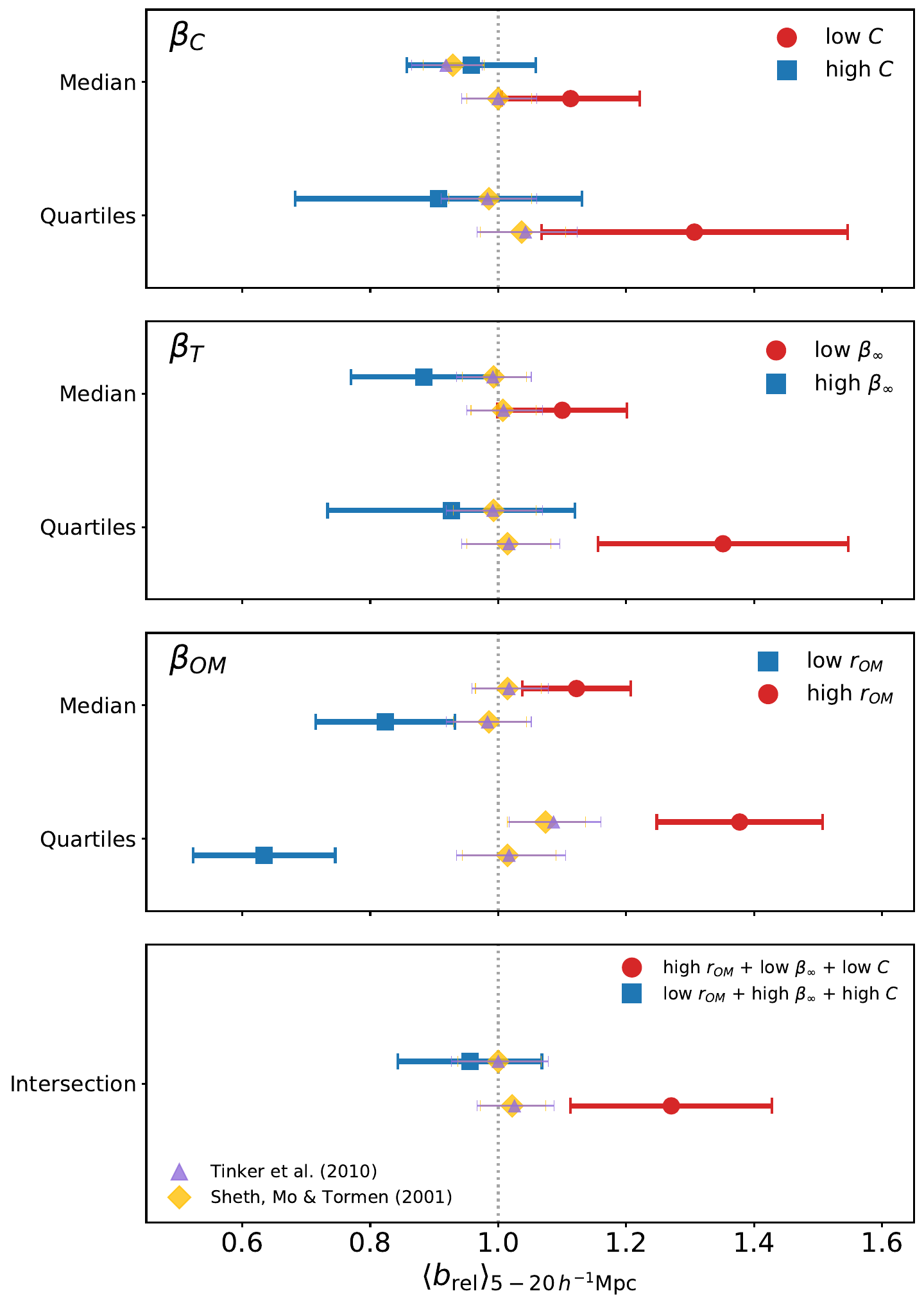}
\label{fig:brel_theory_fixed_rx}
\caption{Same as Fig.~\ref{fig:brel_theory_comparison}, but with all parameters, subsamples, and theoretical predictions recomputed using $r_{200} \equiv r_{200,\rm X}$.}\end{figure}

We compute the projected cross-correlation functions $w_p(r_p)$ and average the relative bias $\langle b_{\rm rel} \rangle$ over scales $5 \le r_p \le 20\,h^{-1}\text{Mpc}$ for the updated low- and high-anisotropy subpopulations. Figure~\ref{fig:brel_theory_fixed_rx} summarises these scale-averaged relative bias measurements alongside the halo mass-only theoretical predictions derived from \citet{Tinker2010} and \citet{Sheth2001}, integrated over the corresponding $r_{200,\rm X}$ distributions.

Crucially, the theoretical mass-only predictions (purple triangles and amber diamonds) remain centred around unity for all subpopulations, reflecting the excellent mass matching achieved using $r_{200,\rm X}$. This supports the conclusion that residual mass variations are
unlikely to fully account for the systematic offset between the
low- and high-anisotropy samples.

These results suggest that the detected clustering segregation is
unlikely to be entirely attributable to the mass--anisotropy degeneracy
inherent to joint dynamical fits. Instead, the signal remains present
when using an independent X-ray mass proxy, supporting its
interpretation as a genuine observational trend.

\end{appendix}
\end{document}